# Hydrodynamic model for laser swelling

N. Bityurin* and N. Sapogova

Institute of Applied Physics of Russian Academy of Sciences, 46 Ul'yanov Street, 603950 Nizhny Novgorod, Russia

## Abstract

The evolution of surface layers of a glassy material heated by a laser pulse above the glass transition temperature and cooled by heat diffusion is considered as the flow of a stretchable viscous fluid. The strong dependence of viscosity on temperature and pressure leads to the appearance of a hump with reduced density. This hydrodynamic model for laser swelling is formulated in general form. We present a 1D solution for laser swelling of a thin glassy polymer film on a strongly thermally conductive substrate for laser pulses long enough that the sound confinement effect can be neglected. It is shown that within this condition the evolution of the film thickness over time can be addressed using a second-order ordinary differential equation. It is also shown that in some cases this equation can be reduced to a first-order differential equation resembling the phenomenological equation of the previously published relaxation model of laser swelling. The main features of the dependence of laser swelling on laser fluence, namely the threshold at low fluencies and saturation at high fluencies, have been clarified allowing for the dependence of viscosity on pressure, which was not taken into account in the previous theoretical studies of laser swelling. Typical regimes of the film thickness evolution are considered and compared to existing experimental data.

*Correspondence: bit@appl.sci-nnov.ru; Tel.: +7 831 4164889

## Introduction

Laser irradiation of glassy dielectrics at fluencies slightly lower than the ablation threshold may result in the appearance of a bump or a hump. If this hump is caused by a decrease in the density of the material, this phenomenon is called laser swelling [1].

This phenomenon has been utilized in laser micro-nano texturing of material surfaces [2,3], including periodic patterning of surfaces using interference techniques [4], some applications in laser printing of micro-sized optical components and micro lens array [5-7], including elaborated structures such as insect-like eyes [8].

Laser swelling has been performed by various lasers with different pulse durations, CW lasers [9, 10], millisecond $CO_2$ laser [11], nanosecond laser [12-17], as well as lasers with pico- [18,19] and femtosecond- [20,21] laser pulses.

Laser irradiation of a material carried out in sound confinement conditions can produce a bump in a way of incomplete spallation [17] due to the generation of a rarefaction wave with negative pressure (tensile stress). Swelling here may originate from the formation of cavitation bubbles [16].

Even if the sound confinement condition is not fulfilled, steam bubbles can arise due to the effervescence of water in biological tissue [22] or residual solvent in polymer samples prepared by casting [23].

Very often, the authors of the mentioned papers prepare glass-like samples by doping them with additional species in order to increase the absorption coefficient at the wavelength of irradiation [24]. Swelling here is often attributed by the authors to generation of gaseous products from the photo/thermochemical decomposition of such compounds [14]. Sometimes gaseous products are formed as a result of matrix decomposition [9]. Gaseous products rupture the matrix, generating voids and bubbles, which lead to swelling.

However, in glassy materials, swelling can occur without the formation of bubbles or voids, and the additional free volume can be uniformly distributed within the hump. This is more useful for optical applications of this phenomenon to avoid parasitic scattering. In this case, swelling is due to the relaxation nature of the glass transition.

When the temperature abruptly increases or decreases just around the glass transition temperature, it takes some time for the density of the material to relax to its equilibrium value. This means that if the laser during a pulse heats the material slightly above the glass transition temperature and thermal diffusion cooling immediately after the pulse occurs faster than the relaxation time, then the density retains its value less than the equilibrium one. It looks like swelling. Glass transition is a rather complex phenomenon [25]. Several attempts have been made to construct a model of laser swelling based on the relaxation nature of this transition [26, 27]. They operate with concepts such as “fictitious temperature” [26], relaxation time [26, 27], etc. These models do not take into account that the relaxation time depends not only on temperature, but also on pressure, as was recently understood [28,29]. Moreover, the above considerations can hardly be generalized to the case of swelling in sound confinement conditions.

In this paper, we present a different approach – a hydrodynamic model of laser swelling. This is a physical model that considers swelling in terms of the generalized Navier-Stokes equation, which addresses the motion of a compressible or, better said, stretchable fluid with a viscosity that depends on temperature and pressure. The strong dependence of viscosity on temperature leads to freezing upon rapid cooling of a thermally stretched fluid in a state with a density lower than the equilibrium value.

The paper is organized as follows. We first formulate the problem in general 3D form, discussing examples of temperature- and pressure-dependent viscosity. Then we focus on the one-dimensional problem of a polymer film on a quartz substrate irradiated by a nanosecond laser pulse, which corresponds to the experimental situation. It is shown that the partial differential equation problem can be replaced by a second-order ordinary differential equation.

Then, using this ordinary equation, we analyze the feature of laser swelling dynamics in comparison with experimental data. Since, according to this model, swelling corresponds to stretching of the material, an important point is to take into account the possibility of the formation of cavitation bubbles. The condition of the absence of bubbles is necessary for the self-consistency of the model considered.

# 1. Formulation of the problem

## 1.1 General consideration

We consider a compressible viscous fluid with a free surface. The system of equations for the flow of this fluid consists of the continuity equation

$$\frac{\partial \rho}{\partial t}+\frac{\partial}{\partial x_i}\left(\rho V_i\right)=0 \qquad (1)$$

and the generalized Navier-Stokes equation (see Eq. 15.5 in [30])

$$\rho\left(\frac{\partial V_i}{\partial t}+V_k\frac{\partial V_i}{\partial x_k}\right)=-\frac{\partial p}{\partial x_i}+\frac{\partial}{\partial x_k}\left\{\eta\left(\frac{\partial V_i}{\partial x_k}+\frac{\partial V_k}{\partial x_i}-\tfrac{2}{3}\delta_{ik}\frac{\partial V_l}{\partial x_l}\right)\right\}+\frac{\partial}{\partial x_i}\left(\zeta\frac{\partial V_l}{\partial x_l}\right). \qquad (2)$$

Here, in Eqs. (1) and (2) $t$ is the time, $\{x_i\}$ are the coordinates, $\rho$ is the fluid density, $V_i$ is the velocity component, $p$ is the pressure, $\eta$ is the shear viscosity, and $\zeta$ is the second, or bulk, viscosity.

This system should be supplemented with boundary conditions. On a fixed solid surface, the boundary condition (see Eq. 15.13 in [30]) is given by

$$V_i = 0 \quad i=1, 2, 3. \tag{3}$$

On a free surface (see Eq. 61.14 in [30]),

$$\left(p-\sigma\left(\frac{1}{R_1}+\frac{1}{R_2}\right)\right)n_i = \sigma'_{ik}n_k + \frac{\partial\sigma}{\partial x_i}, \tag{4}$$

where (see Eq. 15.3 in [30])

$$\sigma'_{ik} = \eta\left(\frac{\partial V_i}{\partial x_k}+\frac{\partial V_k}{\partial x_i}-\tfrac{2}{3}\delta_{ik}\frac{\partial V_l}{\partial x_l}\right)+\zeta\delta_{ik}\frac{\partial V_l}{\partial x_l}, \tag{5}$$

Here, $\sigma$ is the surface tension coefficient, $R_1$ and $R_2$ are the principal radii of curvature at a given point on the surface. The last term on the right-hand side of Eq. (4) addresses the Maranghony phenomenon.

Within this paper, we consider the geometry shown in Fig. 1.

In the one-dimensional case, the system (1)-(2) becomes (here, $\rho=\rho(t,y)$, $V_x=0$, and $V_y=V(t,y)$)

$$\frac{\partial\rho}{\partial t}+\rho\frac{\partial V}{\partial y}+V\frac{\partial\rho}{\partial y}=0, \tag{6}$$

$$\rho\left(\frac{\partial V}{\partial t}+V\frac{\partial V}{\partial y}\right)=-\frac{\partial p}{\partial y}+\frac{\partial}{\partial y}\left(\left(\frac{4}{3}\eta+\zeta\right)\frac{\partial V}{\partial y}\right). \tag{7}$$

With the boundary condition on a plane free surface with coordinate $l(t)$ we have

$$p(l)=\left(\zeta+\frac{4}{3}\eta\right)\left.\frac{\partial V}{\partial y}\right|_{y=l}, \tag{8}$$

where $l$ is the thickness of the film, the initial thickness being $l_0$.

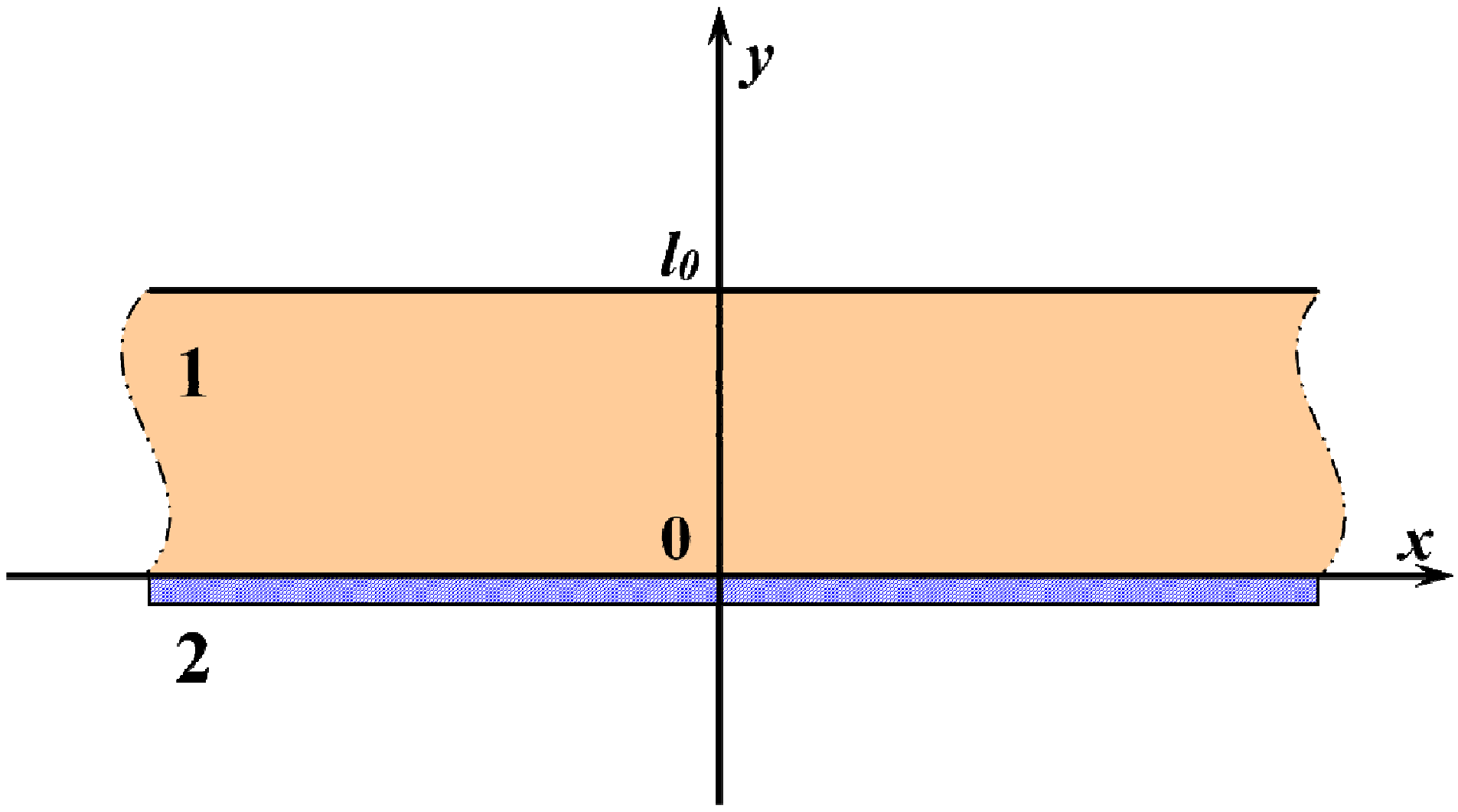

**Fig. 1.** 1– the layer of a glassy medium with thickness $l_0$, 2 – solid substrate. In the two-dimensional case, the variables are functions of the *x* and *y* coordinates. In the one-dimensional case, they are functions of the *y* coordinate.

In the laser swelling phenomenon described by the above equations, the main features are the dependence of viscosity on temperature and pressure. The viscosity value is high enough just at the glass transition point and abruptly decreases with increasing temperature. In what follows, we will use this dependence in the Avramov form (see [28]),

$$\eta = e^{-\gamma}\eta_g \exp\left(\gamma\left(\frac{T_g}{T}\right)^{\alpha}\right), \tag{9}$$

where $T$ is the temperature, $T_g$ is the glass transition temperature, and $\gamma$, $\alpha$ and $\eta_g$ are some constants. It is seen in this expression that the viscosity is determined by the glass transition temperature $T_g$.

At $T$=$T_g$, the viscosity is fixed with the value $\eta_g$=$10^{13.5}$ dPa s = $10^{6.5}$ J s cm$^{-3}$. It is argued in [28] that the temperature at which the viscosity reaches this value can be defined as the glass transition temperature.

This dependence can be compared with the well-known universal dependence for polymers in the Vogel–Fulcher–Tammann form [26, 27, 29]

$$\eta = \eta_0 \exp(\frac{T^*}{T + \Delta T_2 - T_g}) \tag{10}$$

with $\eta_0$ = 1.21938*10$^{-11}$ J s cm$^{-3}$, $T^* \approx 2069K$, and $\Delta T_2 \approx 51.6K$, and $T_g$ = 378 K = *const* for PMMA (see [27]).

Figure 2 shows the correspondence of formulas (10) and (9) with the values $\alpha$≈8.04 and $\gamma$≈31.05.

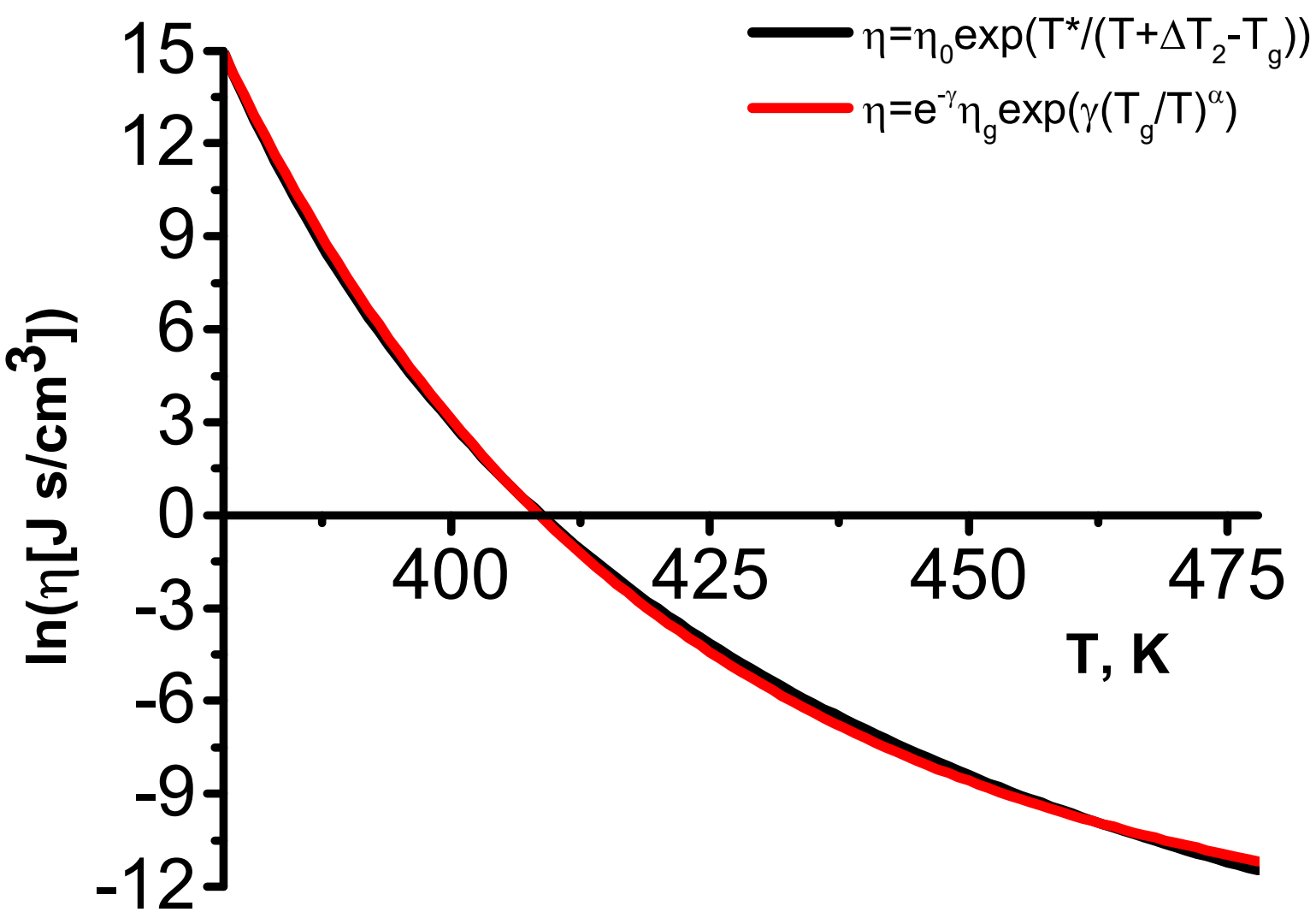

**Fig.2**. Correspondence of temperature dependences of viscosity for functions (10) (black curve) and (9) (red curve) with $\alpha$≈8.04 and $\gamma$≈31.05.

Below we will follow the approximation according to which the bulk viscosity $\xi \approx \eta$ (see [26, 31]).

One of the drawbacks of the relaxation models of laser swelling published in [26, 27] is that the relaxation rate of the material density in them depends only on temperature (see Eq. (10)). Obviously, it should also depend on the excess or lack of free volume in relation to its equilibrium value. This can be accounted for by taking the glass temperature in Eq. (9) as a function of pressure. The exceeding of free volume over equilibrium leads to negative pressure, and the latter leads to a decrease in glass temperature, which, according to Eq. (9), reduces viscosity.

Following [28], we take this dependence in the form

$$T_g(p) = T_{g0}\left(1+\frac{p}{\Pi}\right)^{\frac{\beta}{\alpha}}, \qquad (11)$$

where $\beta$ is another parameter. The parameter $\Pi$ is very important. The smaller $\Pi$, the stronger the dependence of viscosity on variations in free volume. It should be noted that this relation was experimentally obtained mainly for positive pressure; however, it is argued in [29] that there is no singularity at $p$=0.

We take the equation of state in the form

$$p = K\left(\frac{\rho-\rho_g}{\rho} + \bar{\alpha}\left(T - T_g\right)\right), \tag{12}$$

where $K$ is the bulk modulus, $\bar{\alpha}$ is the thermal expansion coefficient, $\rho_g$ is the density at the glass transition point, constant for a specific material.

When the temperature rises above the glass transition temperature, the thermal expansion coefficient is an order of magnitude greater than the coefficient in the glassy state. This increase results from a change in the nature of expansion. Here, thermal expansion is due to an increase in the fraction of free volume. In what follows, we will neglect thermal expansion occurring below the glass transition temperature.

Since the glass transition temperature, according to Eq. (11), is a function of pressure, Eq. (12) has an implicit form.

However, the ratio $p/\Pi$ is usually a small parameter (see [29]), which allows a linear expansion of the right-hand side of Eq.(11),

$$T_g\left(p\right) \cong T_{g0}\left(1 + \frac{\beta}{\alpha}\frac{p}{\Pi}\right). \tag{13}$$

This gives the explicit representation of Eq. (12):

$$p = \frac{K}{1 + \dfrac{KT_{g0}\beta\bar{\alpha}}{\alpha\Pi}}\left(\frac{\rho-\rho_g}{\rho} + \bar{\alpha}\left(T - T_{g0}\right)\right). \tag{14}$$

The above system of equations should be supplemented with equations for the coordinate-time temperature distribution. In the laser swelling problem, the temperature is usually determined by laser heating and heat diffusion.

## 1.2. 1D simplified model

Using one-dimensional hydrodynamic equations, one can arrive at a Lagrange coordinate system (see, e.g., [30]). As the Lagrange coordinate, we fix the initial coordinate of the physical point. Then at time $t$ the position of the point with Lagrange coordinate $a$ is $y(a,t)$, $y(a,0)=a$. The mass conservation gives

$$\int_0^a \rho_0 da = \int_0^{y(a)} \rho(y,t)dy\,, \tag{15}$$

where $\rho_0$ is the initial fluid density.

Differentiating both parts of Eq. (15) with respect to $a$ yields

$$\frac{\partial y\left(a,t\right)}{\partial a} = \frac{\rho_0}{\rho}. \tag{16}$$

Equation (16) and the correspondence

$$\left.\frac{\partial}{\partial t}\right|_y + V\frac{\partial}{\partial y} \to \left.\frac{\partial}{\partial t}\right|_a$$

transform the system (6)-(7) into equations in Lagrange coordinates,

$$\frac{\partial \rho}{\partial t} + \frac{\rho^2}{\rho_0}\frac{\partial V}{\partial a} = 0, \tag{17}$$

$$\rho_0 \frac{\partial V}{\partial t} = -\frac{\partial p}{\partial a} + \frac{1}{\rho_0}\frac{\partial}{\partial a}\left(\rho\left(\frac{4}{3}\eta + \zeta\right)\frac{\partial V}{\partial a}\right), \tag{18}$$

Here the partial derivatives in time are taken at a fixed Lagrange coordinate $a$.

The boundary condition (8) in Lagrange coordinates takes the form

$$p = \frac{\rho}{\rho_0}\left(\zeta + \frac{4}{3}\eta\right)\left.\frac{\partial V}{\partial a}\right|_{a=l_0}. \tag{19}$$

The film thickness is calculated as

$$l(t) = y(a=l_0,t). \tag{20}$$

Below we consider the 1D case (17)-(19) with a simplified temperature distribution in order to analyze the main features of laser swelling within the framework of the model considered.

For the temperature distribution, we will use a rough approximation, taking the temperature constant inside the film (see Fig. 1) and obtaining its time dependence from an equation of Newtonian form,

$$\frac{dT}{dt} = \frac{\alpha_{abs} I(t)}{c_p \rho} - \frac{T - T_{room}}{t_D}, \tag{21}$$

which is approximately valid for sufficiently thin films on a highly thermally conductive substrate, for example, a polymer film on a quartz substrate. Here, $I(t)$ is the time dependence of the laser intensity within the laser pulse, $\alpha_{abs}$ is the absorption coefficient, $c_p$ is the specific heat capacity at constant pressure, $T_{room}$ is the room temperature, and $t_D$ is the heat diffusion time which can be approximated as $t_D = l_0^2/D_T$ with $D_T$ being the heat diffusivity of the film.

The solution of Eq. (21) with the initial condition $T(t=0)=T_{room}$ reads

$$T(t) = \exp\left(-\frac{t}{t_D}\right)\int_0^t \frac{\alpha_{abs} I(t)}{c_p \rho}\exp\left(\frac{t}{t_D}\right)dt + T_{room}. \tag{22}$$

We consider the case where the pulse duration $t_{pulse} << t_D$.

In this case, the temperature will reach the maximum value $T_{max}$ at the moment of time close to the end of the laser pulse

$$T_{\max} = T_{room} + T_0 \quad , \tag{23}$$

$$T_0 \approx \int_{-\infty}^{\infty} \frac{\alpha_{abs} I(t)}{c_p \rho} dt . \quad (24)$$

Below we will characterize laser heating as $T_0$, fixing $T_{room}$ = 293K.

Considering the irradiation of a thin film (thickness about $l_0$~1-2 μm) with nanosecond laser pulses (pulse duration about $t_{pulse}$ ~ 30 ns) leads to the fulfillment of the relation

$$l_0/c << t_{pulse}. \quad (25)$$

Here, c~$10^5$ cm/s is the sound speed. That is, there is no sound confinement.

This means that the pressure inside the film does not vary significantly, which gives the following relation in the Lagrange coordinate system:

$$\frac{\partial p}{\partial a} \approx 0 . \quad (26)$$

Using this constant pressure approximation (26), we can obtain an ordinary differential equation for the time dependence of the film thickness.

Integrating (18) over $a$ from zero to $l_0$ and taking into account the boundary condition (19) yields

$$\rho_0 \int_0^{l_0} \frac{\partial V}{\partial t} da = p - \frac{\rho}{\rho_0} \frac{\partial V}{\partial a} \left( \frac{4}{3}\eta + \zeta \right) \bigg|_{a=0} . \quad (27)$$

Since pressure and temperature do not depend on coordinate within the sample, the density is constant. This means that the speed is a linear function of the coordinate. From Eqs. (16) it follows that

$$V(a) = \frac{1}{l_0} \frac{dl}{dt} a . \quad (28)$$

Bearing in mind that for this case

$$\frac{\rho}{\rho_0} = \frac{l_0}{l} , \quad (29)$$

it follows from Eqs. (27)-(29) that

$$\frac{l_0 \rho_0}{2} \frac{\partial^2 l(t)}{\partial t^2} + \left( \frac{4}{3}\eta + \zeta \right) \frac{1}{l(t)} \frac{\partial l(t)}{\partial t} - \frac{K}{1 + \bar{\alpha} K T_{g0} \dfrac{\beta}{\alpha \Pi}} \left[ \left( 1 - \frac{\rho_g}{\rho_0} \frac{l(t)}{l_0} \right) + \bar{\alpha} \left( T(t) - T_{g0} \right) \right] = 0 . \quad (30)$$

Here, we took into account Eq. (13).

When calculating the viscosities $\eta$ and $\xi$, we use Eqs. (9) and (13), bearing in mind that

$$p(l) = K \frac{1 + \bar{\alpha}\left(T - T_{g0}\right) - \dfrac{\rho_g}{\rho_0} \dfrac{l}{l_0}}{1 + \bar{\alpha} K T_{g0} \dfrac{\beta}{\alpha \Pi}} . \quad (31)$$

All the above equations (17)-(19), as well as Eq. (30), are valid for $T>T_{g0}$. Thus, during heating, we neglect the temperature expansion of the film if $T<T_{g0}$ and start calculations with $T=T_{g0}$. During cooling, we finish the calculations at $T=T_{g0}$. At this temperature, the viscosity is so high that the film thickness is actually fixed. This means that we solve systems of equations with the 'initial' condition $l=l_0$ at $T=T_{g0}$ and $\rho_0=\rho_g$.

Together with Eq. (30) we consider two reduced equations. Equation (30) describes a nonlinear strongly damped oscillator.

Similarly to the harmonic oscillator, if the inequality

$$\frac{l_0}{\overline{\eta}}\sqrt{\frac{\rho_0\overline{K}}{2}} \ll 1 \tag{32}$$

is satisfied, then the reduced equation can be obtained from Eq. (30), neglecting the first term with the second time derivative,

$$\overline{\eta}\frac{1}{l(t)}\frac{\partial l(t)}{\partial t} - \overline{K}\left[\left(1-\frac{l(t)}{l_0}\right)+\overline{\alpha}\left(T(t)-T_{g0}\right)\right]=0\,. \tag{33}$$

Here, in Eqs. (32) and (33), we have $\overline{\eta}=4/3\eta+\zeta$ and

$$\overline{K}=\frac{K}{1+\overline{\alpha}KT_{g0}\dfrac{\beta}{\alpha\Pi}}.$$

It follows from inequality (32) that Eq. (33) is valid for a fairly small film thickness.
Eq. (30) corresponds to the quasi-stationary approximation where pressure is balanced by the viscous forces.

It is seen from Eq. (33) that if instead of $l$ here we consider the variable $l/l_0$, then the initial film thickness $l_0$ enters the equation only within the limits of this variable. This is not the case in Eq. (30). Here, the parameter $l_0^2\rho_0/2$ appears exactly on the term with the second derivative.

The second reduced equation corresponds to zero pressure and is equal to the equilibrium value of the film thickness at a given temperature.

$$\frac{l(t)}{l_0}=1+\overline{\alpha}\left(T(t)-T_{g0}\right). \tag{34}$$

With all the below calculations, the numerical solutions of the system (17)-(19) coincide with the solution of Eq. (30) with great accuracy.

## 2. Results and Discussions

Consider typical modes of behavior of the irradiated film. We compare the calculation results with experimental data published in [15]. Here, the time evolution of the polymer (PMMA) film thickness has been studied. A film with an initial thickness of 2 μm was deposited on a quartz substrate and irradiated with a KrF (248 nm, 30 ns) laser pulse. The film thickness evolution was monitored using the interferometer technique.

Experimental data, as the authors note, are somewhat controversial. However, the observed thickness evolution is typical. We address these data using ~~typical~~ parameter values for PMMA and following [27] using the maximum temperature increment $T_0$ as a fitting parameter.

For the numerical solution we use the following values of parameters: $t_D = 4*10^{-5}$ s, $T_{room} =$ 293 K, $l_0$= 2 $10^{-4}$ cm, $\rho_0 = \rho_g = 1.19$ g/cm$^3$, $\eta/\zeta$=1, $K = 10^3$ J/cm$^3$, $\tilde{\alpha} = 10^{-3}$ K$^{-1}$, $T_{g0} = 378$ K, $\alpha =$ 8.04, $\eta_g = 3.16*10^6$ J s/cm$^3$, $\gamma = 31.05$, and $\beta = 2.34$.

The dynamics of laser swelling depends on $T_0$ (see Eq. (24)). If the maximum temperature $T_{max}$ (see (23)) is higher, but close to the glass transition temperature, then the viscosity is quite high even at $T$=$T_{max}$. This means that the thickness relaxation is much slower than the pulse duration. At the end of the laser pulse the thickness retains its initial value. This leads to an initial increase in pressure followed by a slow increase in thickness. The increase in thickness results in a decrease in pressure. During cooling, the thickness lags behind its equilibrium value, reducing the pressure to a negative value. This is illustrated by Figs. 3a and 3b.

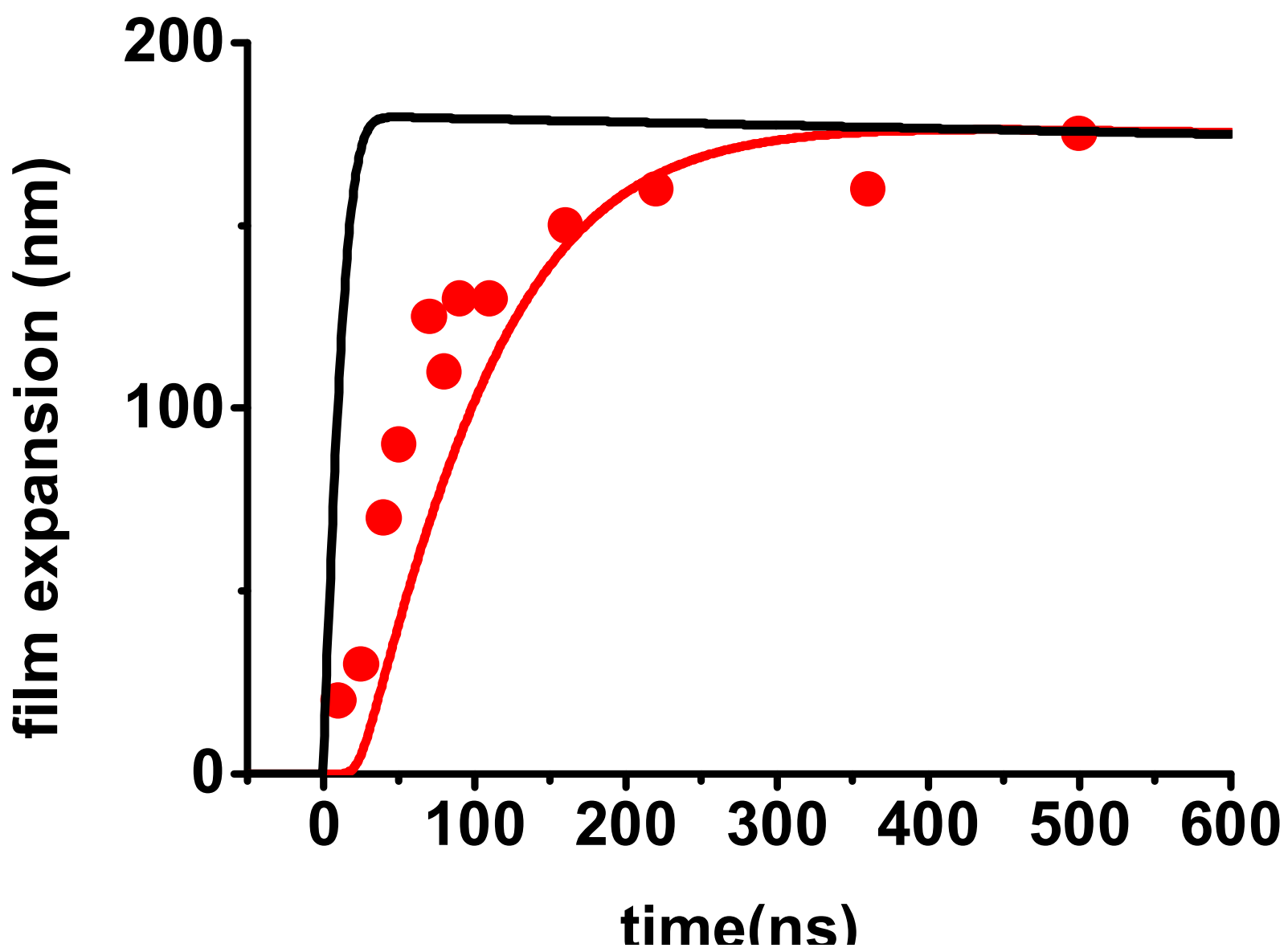

**Fig. 3a.** Dependence of film expansion on time. The time $t$=0 corresponds to the maximum of the pulse. The circles are experimental points [15]. The red solid curve is the solution of Eq. (30) coinciding with the solutions of Eq. (33) and solution of the system (17)-(19). The black curve is the solution of Eq. (34), and $\Pi$=2500 J/cm$^3$. Pulse duration $t_{pulse}$ =30ns, $T_0$=175K.

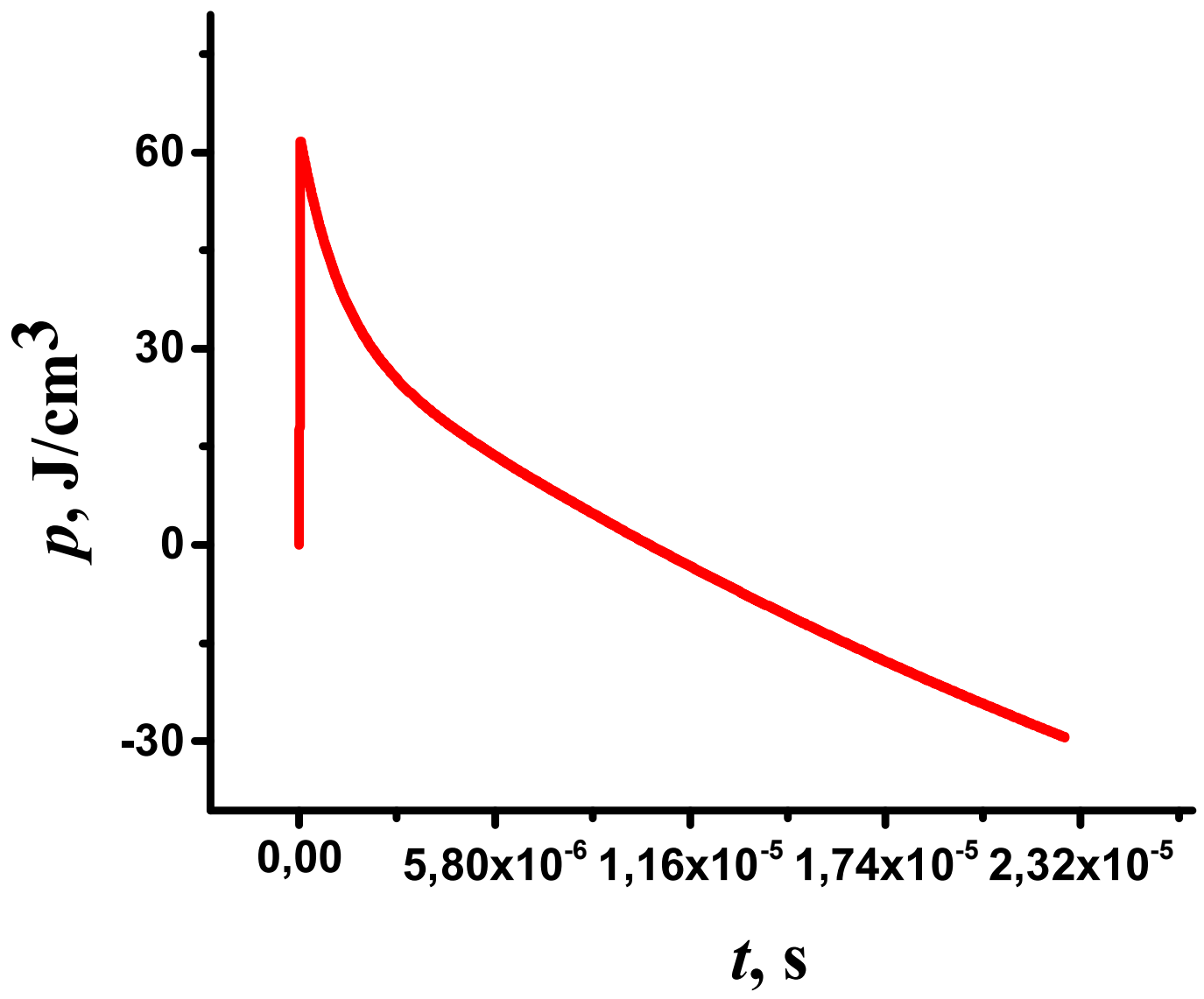

**Fig. 3b.** Time dependence of pressure $p(t)$ for heating above room temperature at $T_0$=150K for $\Pi$=2500 J/cm$^3$ and $t_D$ = 4*10$^{-5}$s.

If the maximum temperature is high enough (see Fig. 4), then the viscosity value after the pulse is relatively small, i.e., until significant cooling the thickness behaves in such a way that the effect of viscosity can be neglected. For a pulse duration of 30 ns, the thickness approaches the equilibrium value at the end of heating. During cooling, the thickness first follows the equilibrium value considered by Eq. (34), keeping the pressure close to zero (Eq. (34) corresponds to zero pressure). Gradually, a decrease in temperature leads to an increase in viscosity. The thickness starts to vary more slowly than in the equilibrium state (Eq. (34)), approaching the swelling limit. This behavior is illustrated by Fig. 4a for short times and Fig. 4b for times comparable to the cooling time.

The figures show that the solution of Eq. (30) can match well with the experimental curve. The time period of the first 100 ns near the laser pulse is the exception. It seems that oscillations are present here. Of course, these oscillations cannot be addressed by Eq. (33). If we assume an effective pulse duration of 10 ns, the solution of Eq. (30) shows oscillations (see the black curve in Fig. 4a), but the period of these oscillations is too small. We were able to fit the experimental curves more appropriately by only assuming the dependence of the bulk modulus $K$ on free volume or on the film thickness $K=K_0\exp(-\Omega(l-l_0)/l_0)$ at $\Omega$=20 (see the cyan curve in Fig. 4a). Importantly, a comparison of the solutions of Eqs. (17)-(19) and Eq. (30) at short times shows a good agreement. This means that the film oscillates as a single unit.

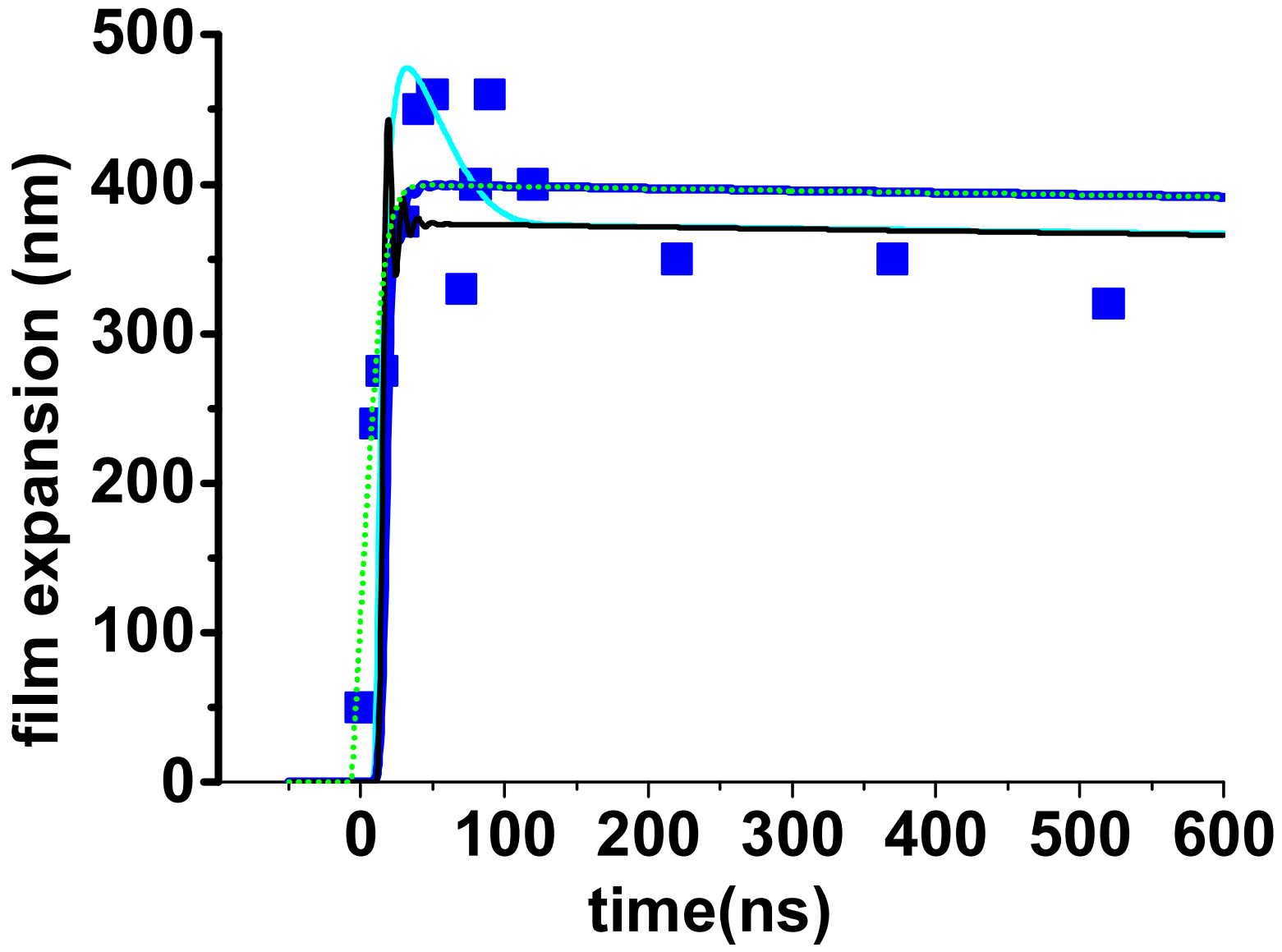

**Fig. 4a.** Time dependence of the film expansion. The time $t$=0 corresponds to the pulse maximum.

Squares are experimental points [15].

Pulse duration $t_{pulse}$ = 30ns.

The blue solid line is the solution of Eq. (30) coinciding with the solutions of the system (17)-(19). $\Pi$=2500J/cm$^3$. The green dotted curve is the solution of Eq. (34). $T_0$=285K.

Pulse duration $t_{pulse}$ = 10ns.

The solid black line is the solution of Eq. (30), coinciding with the solutions of the system (17)-(19). Π=2500J/cm$^3$. $T_0$=272K. The cyan curve is the solution of Eq. (30), coinciding with the solutions of the system (17)- (19). Π=2500J/cm$^3$. $K=K_0\exp(-\Omega(l-l_0)/l_0)$, with $K_0$ = 10$^3$ J/cm$^3$ and $\Omega$=20.

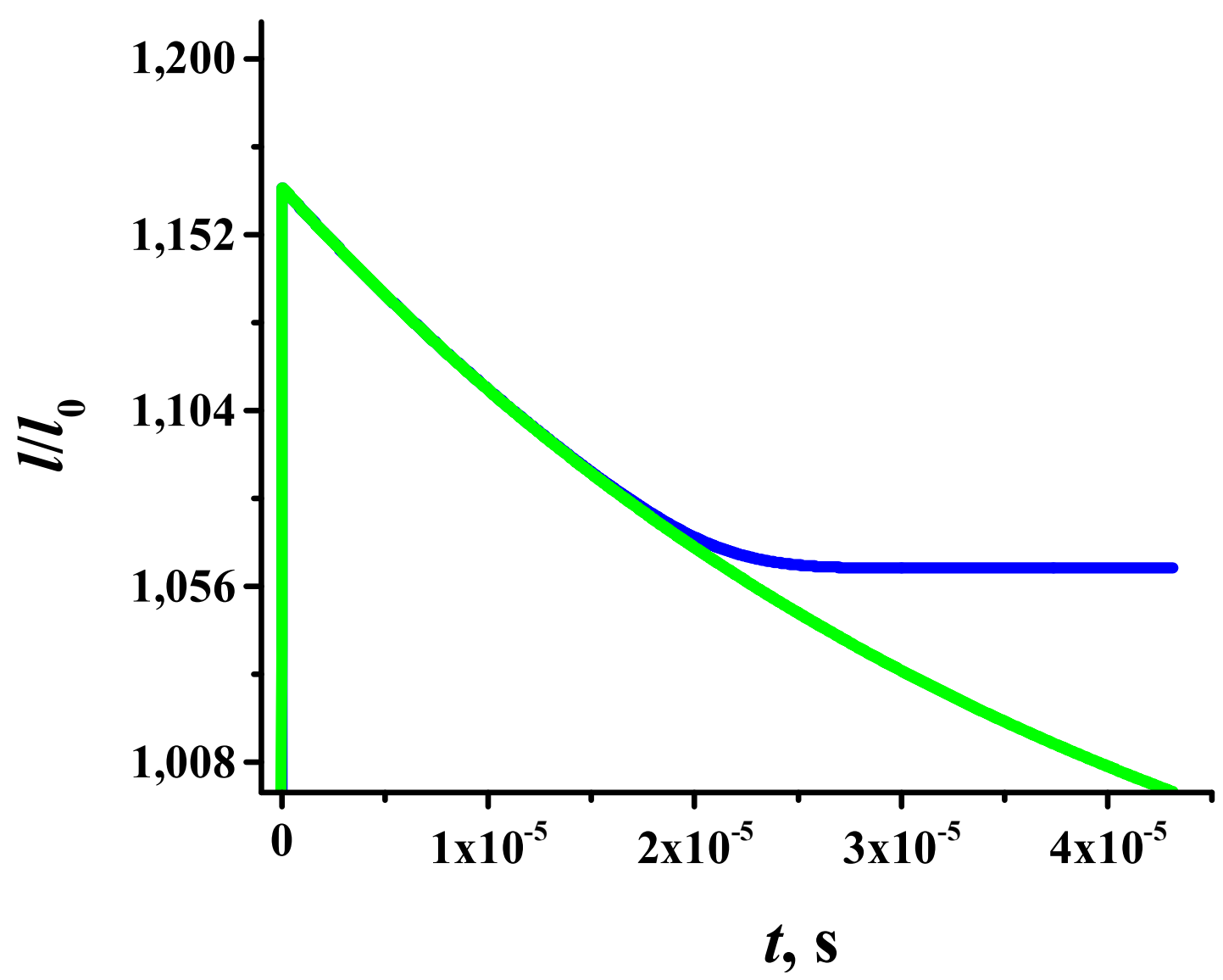

**Fig. 4b.** Dependence $l(t)/l_0$ for heating above room temperature at $T_0$=250K for $\Pi$=2500J/cm$^3$ and $t_D$ =4*10$^{-5}$s. The blue line corresponds to laser irradiation by the Gaussian beam of pulse

duration 30 ns, the solution of Eq. (30), and the green solid line, to laser irradiation by the Gaussian beam of pulse duration 30 ns, calculated by Eq. (34).

The dependence of the final thickness of the film normalized by the initial one after laser heating followed by cooling on the maximum increment temperature $T_0$ is shown in Fig. 5

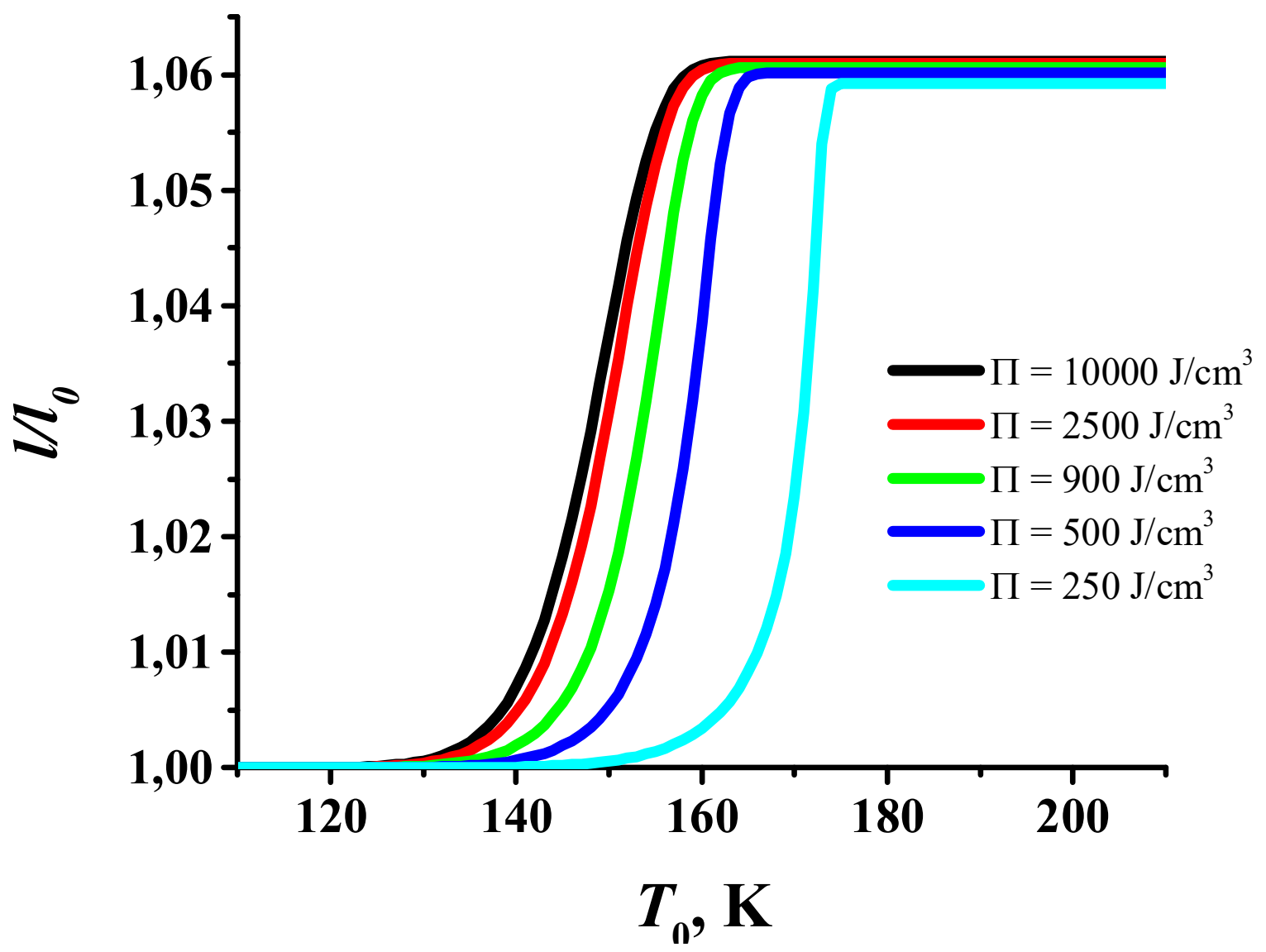

**Fig.** 5. Dependence of the normalized final thickness on the heating temperature for various values of the parameter *П*. $T_g$=378K and $t_D$ =4*10$^{-5}$s.

It is seen in Fig. 5 that the swelling effect is a threshold one. The threshold increases with decreasing parameter *П*. The second feature of this figure is the existence of the limiting value of the swelling. By increasing the maximum temperature, it is not possible to obtain swelling greater than some limiting value.

The explanation of this limitation is quite natural. Indeed, at a sufficiently high temperature, the viscosity is small enough that the film thickness after pulse cooling corresponds to the solution of Eq. (34). At the temperature at which the viscosity starts to be significant, the thickness starts to lag behind the equilibrium thickness, and the value of this temperature does not depend on the initial heating. This is illustrated by Fig. 4b.

Both the threshold temperature and the plateau temperature as functions of the parameter *П* are shown in Fig.6.

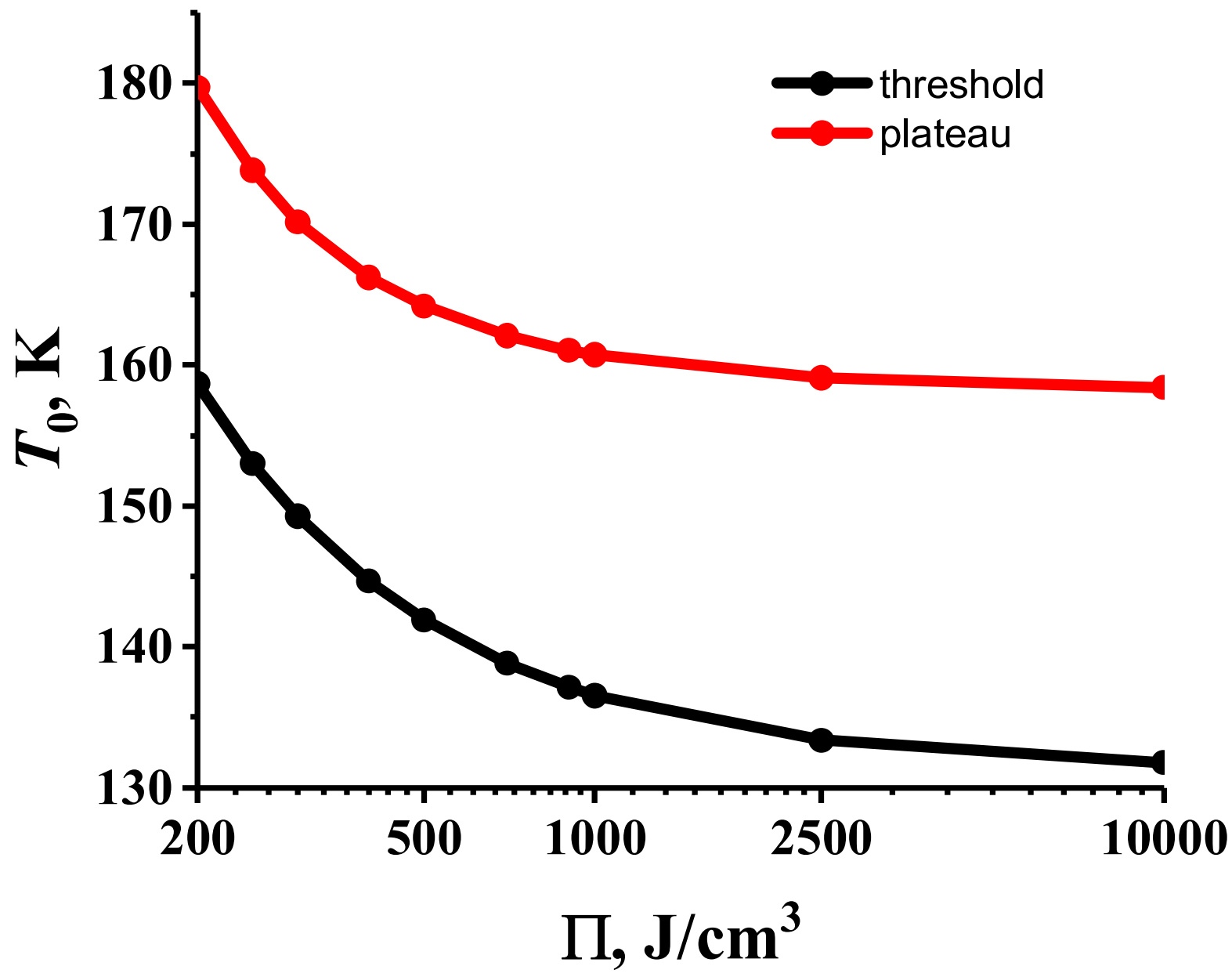

**Fig. 6.** Dependences of the threshold heating temperature for swelling (black line) and the plateau temperature at which the maximum swelling is reached (red line) on the parameter $\Pi$ for the parameters $T_g$=378K and $t_D$ =4*10$^{-5}$s.

Figures 5 and 6 show that the dependence of viscosity on non-equilibrium free volume (i.e., dependence of the glass transition temperature on pressure) prevents swelling.

The maximum value of residual swelling depends on the cooling time. An example of this dependence is shown in Fig. 7.

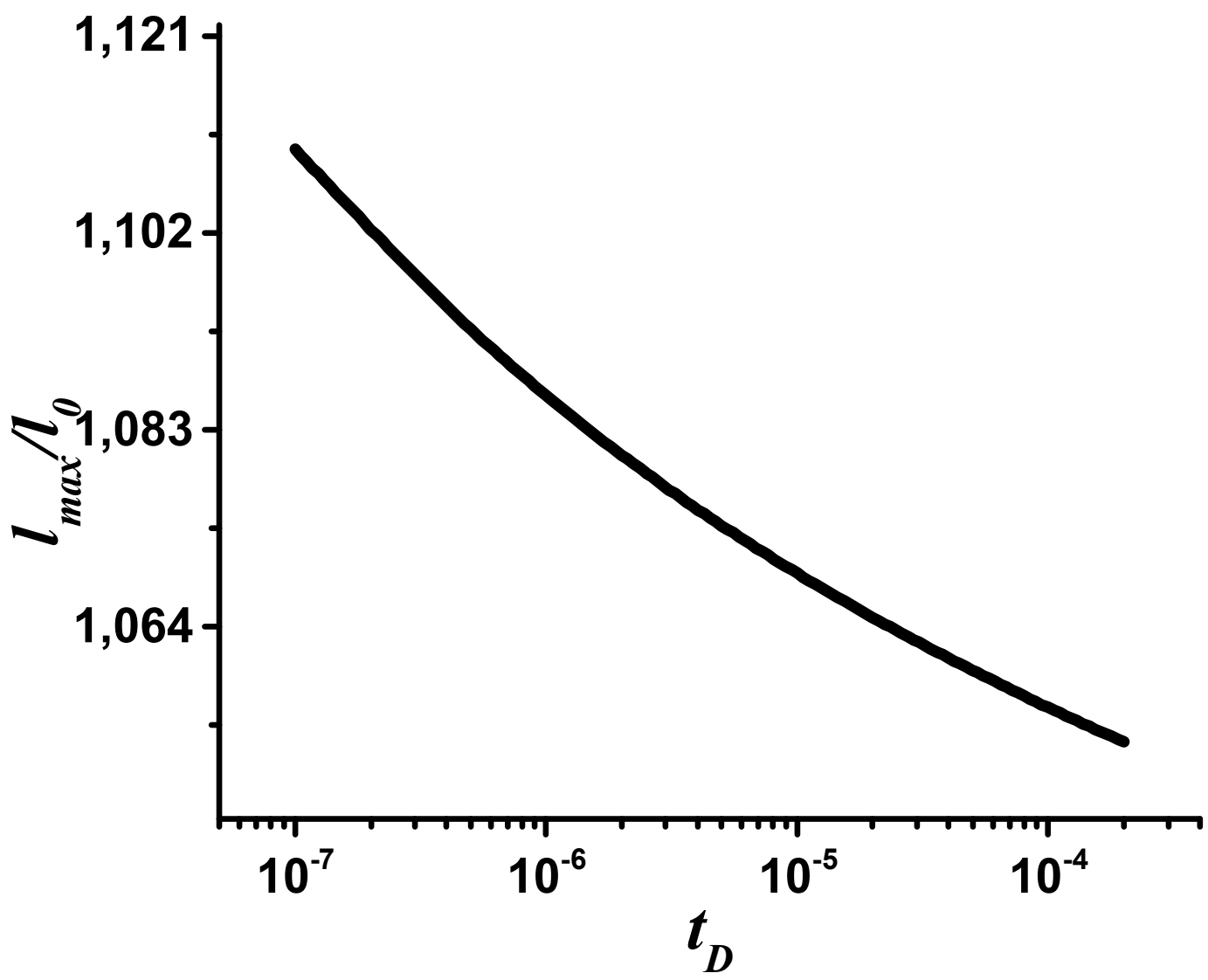

**Fig. 7.** Dependence of the maximum residual thickness normalized to the initial value on the cooling time.

Negative pressure due to swelling can lead to the generation of cavitation bubbles. This process, according to [33], can be considered as a first-order phase transition. Bubble generation is characterized by the value of the critical radius $r_{cr} = -2\sigma/p$ and the nucleation rate (see [16] for references)

$$J = n_0 \frac{\sigma}{\eta} \sqrt{\frac{\sigma}{kT}} \exp\left( -\frac{16\pi\sigma^3}{3kTp^2} \right). \quad (35)$$

This is the rate of generation of the bubbles with radius greater than the critical value.

The surface tension coefficient can be addressed by the Guggenheim expression $\sigma = \sigma_0(1-T/T_c)^{11/9}$ with $\sigma_0 \sim 36.5\ 10^{-7}$ J/cm$^2$, and $T_c \sim 1000$K (see [16] for references).

Once generated, the subcritical bubble within the highly viscous fluid will grow at the rate

$$\frac{dr}{dt} = -\frac{p(r - r_{cr})}{4\eta}.$$

Calculations using the above formulas show that the nucleation rate is very low. An example of the calculation results can be seen in Fig. 8. That is, the generation of bubbles in the considered phenomenon can be neglected. This is in sharp contrast to the case considered in [16], where bubbles are generated by a rarefaction wave. There, relation (25) is not valid, bubbles are formed at the end of the laser pulse when the viscosity is relatively low, the negative pressure is high, and the bubbles have time to grow before effective cooling, resulting in bubble preservation due to high viscosity. In the case considered in the present paper, relation (25) is valid. Here, at low viscosity, the pressure is small and an increase in negative pressure is accompanied by an increase in viscosity, thus preventing the generation and growth of bubbles.

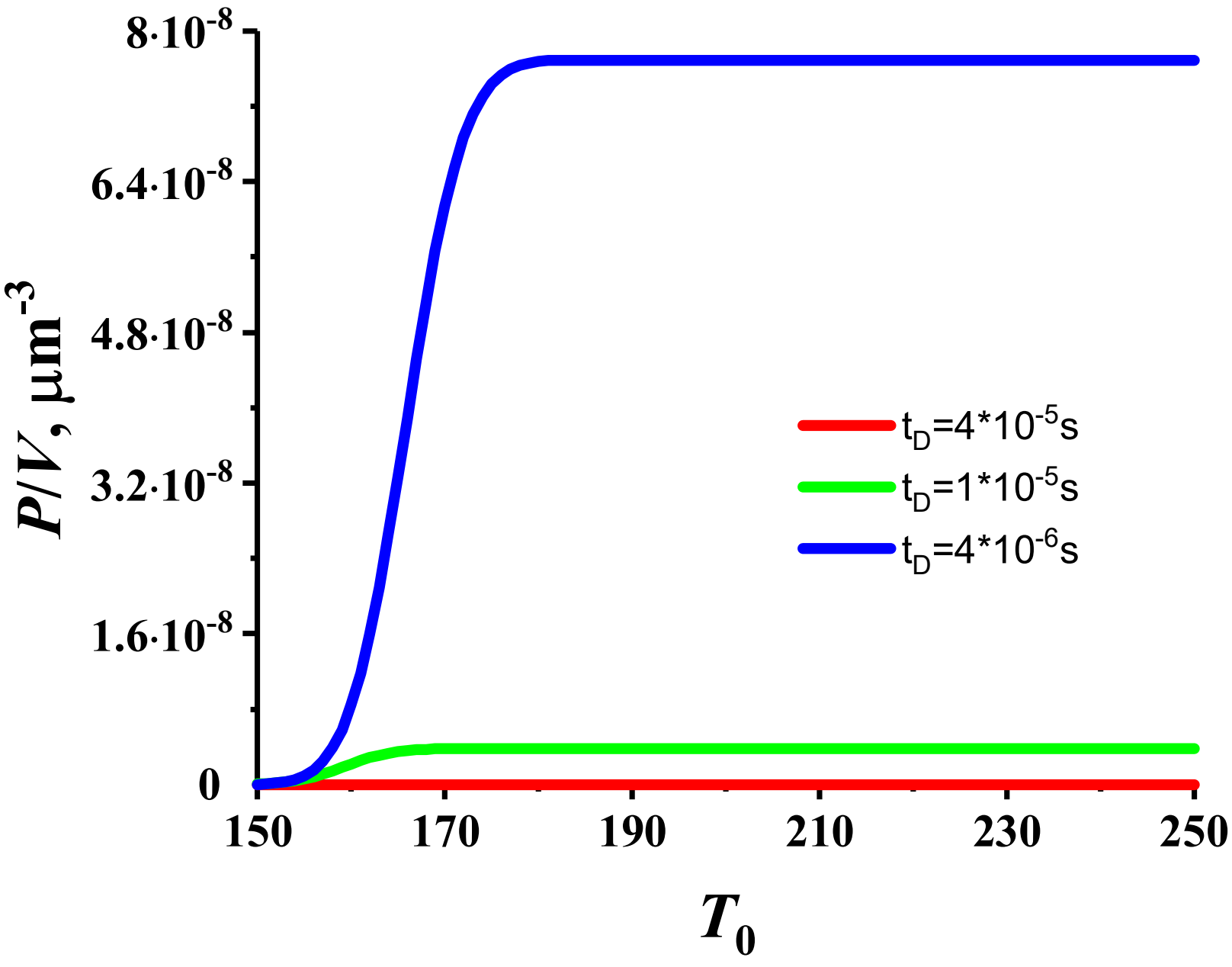

**Fig. 8.** Probability density $P = \int_0^\infty J dt$ (see (35)) as a function of the temperature increment $T_0$ for different values of cooling time $t_D$. $n_0 = 7.7*10^{20}$ cm$^{-3}$ [16].

The curvature of the surface in accordance with the boundary condition (4) leads to an additional pressure $p_{addit} = 2\sigma/R$. Here, $R$ is the curvature radius.

According to the above consideration (see e. g. Fig. 3b), the actual pressure in the sample is more than $p$~1J/cm$^3$. With $\sigma$~36.5 10$^{-7}$ J/cm$^2$, the radius $R$ at which the additional pressure would be of the order of such a pressure should be $R$~100 nm. That is, the nano swelling requires special consideration.

Above we have considered the swelling of a thin layer that is uniformly heated and uniformly cooled. Two main effects were found. We pointed out that there the swelling has a threshold and that at sufficiently strong heating there is a saturation of the value of the residual expansion of the film. If we consider a bulk sample heated by laser radiation with the distribution of absorbed energy governed by the Beer-Lambert law, then the threshold nature of the effect should be preserved, whereas the saturation effect is not obvious and requires special consideration. Here, saturation should occur in a particular layer, but the number of heated layers should increase with increasing fluence due to heat diffusion.

The 1D consideration of laser swelling presented in this paper can be generalized to 2D and 3D problems. This will make it possible to consider more complex phenomena than hump growth, in particular sombrero-like structures [13], etc. Phenomena related to material stretching under the action of tensile stresses generated by a rarefaction wave, including the generation and growth of bubbles, can also be considered within the hydrodynamic model.

We have considered the simplest version of the equation of state (12). This equation can be supplemented with terms that take into account the additional pressure created by gaseous substances dissolved in the material.

## 3. Conclusions

We have considered one of the mechanisms of the laser swelling phenomenon due to laser heating of glass material above the glass transition temperature and rapid cooling due to heat diffusion. Laser heating takes place in a regime that excludes the generation of a rarefaction wave.

To address the laser swelling effect, we present a model considering the motion of a stretchable viscous fluid. Here, the viscosity of the fluid strongly depends on temperature and excess free volume, modeling the situation known for glassy solids above the glass transition temperature.

Here, laser swelling results from thermal expansion with incomplete relaxation to the equilibrium value after cooling, since the relaxation process is hindered by a sharp increase in the viscosity value with decreasing temperature.

This paper considers the one-dimensional case allowing a transition to Lagrange coordinates and a uniform temperature distribution in a thin glassy layer deposited on a solid substrate. The layer is heated by a Gaussian laser pulse to a maximum temperature greater than the glass transition temperature, which after the pulse relaxes exponentially to a room temperature smaller than the glass transition temperature. This is close to the experimental situation discussed in the paper [15], which studied the laser swelling of a polymer film on a quartz substrate.

Within the framework of the above consideration under the assumption that the pressure inside the film is coordinate independent, the system of equations can be reduced to a second-order ordinary differential equation for the film thickness. The equation describes a non-autonomous nonlinear strongly damped oscillator. Analysis of the solutions of the full and reduced equations shows that the solution of the reduced equation matches well with the solution of the complete system. Moreover, it is shown that the problem can be further simplified by considering the first-order ordinary equation, neglecting the term with the second time derivative. It corresponds to the quasi-stationary approximation where pressure is balanced by the viscous forces.

Consideration of the above mathematical problem shows the actual threshold nature of laser swelling observed in the experiments.

With a sufficiently high laser heating, the film thickness rapidly approaches the equilibrium thickness value for the maximum temperature and then the thickness relaxes with decreasing temperature upon cooling, arriving at some limiting value. This limiting or plateau value is almost independent of the maximum temperature.

We have analyzed the role of the dependence of viscosity on excess free volume, which is formalized as the dependence of glass temperature on pressure. This dependence is characterized by the parameter $\Pi$. The smaller the value of $\Pi$, the higher the dependence of the viscosity on the fraction of free volume. We show that decreasing the value of $\Pi$ leads to an increase in the swelling threshold and also to an increase in the plateau temperature. This reduces slightly the thickness of the plateau, but this effect is not very strong.

In the considered mechanism, laser swelling is usually not accompanied by bubble generation, unlike laser swelling related to a negative pressure refraction wave, where the swelling looks like incomplete spallation. The considered mechanism is very useful for applications where swelling is used to create micro-lens systems. The presence of bubbles is undesirable here.

## Acknowledgments

This work was supported by the Russian Science Foundation under project No. 22-19-00322.